\documentclass[twocolumn,showpacs,preprintnumbers,amsmath,amssymb]{revtex4}
\usepackage{graphicx}
\usepackage{dcolumn}
\usepackage{bm}
\begin{document}
\title{Decoherence of two-electron spin states in quantum dots}
\author{D. D. Bhaktavatsala Rao, V. Ravishankar and V. Subrahmanyam}
\affiliation{Department of Physics, Indian Institute of Technology Kanpur, INDIA.}
\date{\today}
\begin{abstract}
The time evolution of spin states of two electrons interacting with a  nuclear
spin bath in a quantum dot system is studied.
The hyperfine interaction between the electrons and the nuclear spins is modeled by an isotropic Heisenberg interaction,
and the interaction between the electron spins by Heisenberg exchange.
Depending on the extent of the overlap between the spatial wave functions of the
electrons, there are two physically different cases, namely the two qubits either interact with
the same set of nuclear spins or they see different nuclear spin environment. In the two cases, the decoherence of the
two-qubit state is studied analytically. 
We have identified a class of two-qubit states which have a rich dynamics when the exchange
interaction between the qubits becomes large in comparison to the hyperfine interaction strengths. The decoherence time
scale is determined as a function of the bath-spin distribution and the 
polarizations of the initial two-qubit state. 
States with large decoherence times are identified by performing a minimization over all the two-qubit pure states.

\end{abstract} 
\pacs{Valid PACS appear here}
\maketitle
\section{Introduction}
Decoherence in quantum systems is a major obstacle for quantum computation and
information processing, causing errors during the implementation of various gate
operations.
Decoherence is well-studied in single qubit systems
which are realized by pseudo two-level systems in Superconducting Quantum Interference Devices (SQUIDS) \cite{squid}, Nuclear Magnetic \text{\text{Re}}sonance (NMR) \cite{slloyd}, and Quantum Dot \cite{qdqc} systems.
Although single qubits form the fundamental unit for quantum computing, multi-qubit systems are required to perform non-trivial gate operations and implementing quantum algorithms \cite{sguide}. The study of decoherence in multi-qubit systems
is, therefore, quite important. 
It is a more involved task here,
as one has to look at the effects of environment on both the loss of coherence and the 
entanglement of the initial state.
As entanglement is believed to exponentially speed up the implementation of quantum algorithms \cite{ekert},
the time scales for the entanglement loss are as important as the decoherence
time scales. 


In this work we study the spin dynamics of two electrons confined in a quantum dot geometry.
That is, we inject a
pair of electrons (in a definite spin state) into a confined region of nuclear spin bath. 
The two electrons are coupled to each other by an exchange interaction, and to the nuclear spins through a 
contact hyperfine interaction.
The exchange interaction strength $J$ is a measure of the overlap of the spatial wave functions of the two 
electrons (qubits). There are two distinct physical cases we have to consider. The spatial wave functions of the
electrons can have a strong overlap,
indicating a large value of $J$. Also both electrons interact with the same set of neighboring nuclear spins here,
which indicates a common nuclear bath for the two-qubit system. On the other hand, when the two electrons are physically
apart, the overlap is quite small indicating a smaller value of $J$.  Here, each electron interact with a different set
of neighboring nuclear spins, and thus implying a different spin bath for each electron.
Since $J$ can be controlled by gate voltages one can tune from one regime to the other. \text{\text{Re}}cent experiments
 \cite{petta1, petta2} have shown a strong dependence of decoherence time scales
on the exchange interaction between the qubits. In a semi-classical treatment given by Taylor et al \cite{petta3},
the interaction of the qubits with the nuclear spins is replaced by an effective
magnetic field. The decoherence time scales are found by performing an ensemble average of
the spin expectation values over different configurations of the magnetic field. 
A recent review by Hanson et al \cite{hanson} summarizes the results of experimental procedures employed in studying the spins in few-electron quantum dots. 
Gywat et al \cite{oliver} have studied the evolution of qubits interacting with an 
off-resonant cavity (instead of a nuclear spin bath). Decoherence of two electrons, interacting with different nuclear
baths,  in the presence of a magnetic field has been studied by Zhang et al \cite{zhang}. The decay of the singlet-triplet
correlator in a double dot system was studied by Coish and Loss \cite{loss}.
The evolution of a central two-qubit system interacting with a spin bath is studied by several authors \cite{hamdouni, dobro}.
A bath-induced entanglement between initially unentangled qubits has been studied earlier \cite{benatti, braun}.
It was shown by Storcz et al \cite{storcz} that the errors during gate operations involving coupled solid state qubits
can be reduced when both the qubits are coupled to a common bath at low temperatures.
Effects of decoherence on the transfer of quantum states along a spin chain are studied in
\cite{sbose}, where the authors have considered separate spin bath environments for each qubit.

In our earlier work \cite{durga,durga2} we have shown that the hyperfine interaction between a qubit
and the nuclear spin in quantum dots can be mapped to an effective isotropic Heisenberg interaction between the qubit and
the neighboring nuclei. Such an approximation is valid over time scales where the contribution of
the sub-dominant terms in the Hamiltonian are weak. Under the same approximation, a model Hamiltonian in the present case 
of two qubits interacting with the nuclear spins can be written as
\begin{eqnarray}
\label{solvham}
\mathcal{H}=(K_A\vec{S}_A\cdot\vec{I}_{\mathcal{E}_A}+K_B\vec{S}_B\cdot\vec{I}_{\mathcal{E}_B}) + J\vec{S}_A\cdot\vec{S}_B,
\end{eqnarray}
where the qubit spins $\vec{S}_{A}, \vec{S}_B$ couple to the total nuclear spins $\vec{I}_{\mathcal{E}_A}$,  $\vec{I}_{\mathcal{E}_B}$, with
respective coupling strengths $K_A$ and $K_B$. It should be noted that $\vec{I}_{\mathcal{E}_A}$ obtained by summing over
all the nuclear spins that interact with the qubit $A$ through a contact hyperfine interaction. 
The exchange interaction between the two qubits is given by the last
term in Eq.\ref{solvham}. As we argued earlier, depending on the exchange interaction strength $J$, we will consider
two physically different situations, viz. (i) a common bath for the qubits and
$\vec{I}_{\mathcal{E}_A}=\vec{I}_{\mathcal{E}_B}$ (when $J$ is large, i.e.
the spatial wave functions of the qubits strongly overlap) (ii) different baths for the two qubits and
$\vec{I}_{\mathcal{E}_A}\ne \vec{I}_{\mathcal{E}_B}$ (when $J$ is small, i.e. the two qubits are physically far apart).
For a GaAs quantum dot, typical values of $K_A$ and $K_B$ \cite{petta3} are $\Large{O} (10^{-8} eV)$.
Depending on whether the bath is common or different, the exchange interaction strength $J$ can take values
from a few $100 \mu eV$ to a few $10 neV$. The number of nuclei $N$ with which the qubits are strongly interacting, can vary too. When the baths are isolated, $N \sim 10^3$ \cite{lee}. On the other hand, when the overlap between the electrons is very large, they together see a larger nuclear bath with
$N \sim 10^5$.

\subsection{The initial state}
We take the initial qubit-bath state to be a direct product,
$\rho(0) = \rho_{AB}(0) \otimes \rho_{\mathcal{E}}(0)$.
$\rho_{AB}$ can be represented as
\begin{equation}
\label{tqs}
\rho_{AB}={\hat{\mathcal{I}}\over 4} + {1\over 2}\vec{P}_A\cdot\vec{S}_A + {1\over 2}\vec{P}_B\cdot\vec{S}_B + \sum_{m,n=1}^{3}\Pi^{mn}S^m_{A}S^n_{B},
\end{equation}
where the vector polarizations are given by $\vec{P}_{A,B} \equiv 2Tr[\rho_{AB}\vec{S}_{A,B}]$, and the
cartesian components of the tensor polarization are $\Pi^{mn}\equiv 4Tr[\rho_{AB}S^m_AS^n_B]$.
We note that for a pure state ($\rho^2_{AB}=\rho_{AB}$), we have $P_A=P_B\le 1$, and 
$P^2_A + P^2_B + \sum_{mn}(\Pi^{mn})^2 = 3$. It should be noted that 
pure states have a non-vanishing tensor polarization. 
If $\rho_{AB}$ is also maximally entangled, then we have$P_A=P_B=0$, implying maximal tensor polarization strengths.
On the other hand,
for direct product pure states, we have $\Pi^{mn} = P^m_AP^n_B$. $\rho_{AB}$ for a maximally-entangled
state $\rho^E_{AB}(0) = \frac{1}{2}[| \uparrow \downarrow \rangle - | \downarrow \uparrow \rangle][ \langle \uparrow \downarrow | - \langle \downarrow \uparrow |] $, and a direct product state $\rho^U_{AB}(0) = | \uparrow \downarrow \rangle \langle \uparrow \downarrow |$ can be written in terms of the various polarizations as 
\begin{eqnarray}
\label{examppols}
\rho^E_{AB}&=&\frac{1}{4}[\hat{ \mathcal{I}} - 4(S^x_{A}S^x_{B}+S^y_{A}S^y_{B}+S^z_{A}S^z_{B})], \nonumber \\
\rho^U_{AB}&=&\frac{1}{4}[\hat{\mathcal{I}} + 2({S}^z_A -{S}^z_B) -4S^z_{A}S^z_{B}]. 
\end{eqnarray}

We now model the initial state of the bath following the approach taken by
Rao et al \cite{durga}. We write the state of the bath
as an incoherent superposition of states labeled by the 
bath spin $I_{\mathcal{E}}$, with weights $\lambda_{I_{\mathcal{E}}}$, $\rho_{\mathcal{E}}(0)= \sum \lambda_{I_{\mathcal{E}}} \rho_{I_{\mathcal{E}}}(0)$. In this study
all $\rho_{I_{\mathcal{E}}}(0)$ will be taken to be unpolarized (multiple of identity). The weights $\lambda_{I_{\mathcal{E}}}$ are however, free parameters. 
This is not too-restrictive an assumption since we have shown in \cite{durga} that higher-order polarizations in each
spin sector of the bath give small corrections to the decoherence time scales. 
Depending on whether the bath spins interact ferromagnetically or antiferromagnetically $\langle \hat{I}^2_{\mathcal{E}} \rangle$
can be large ($\sim N^2$) or small (close to zero), where $N$ is the number of nuclear spins.
For a completely unpolarized bath with $\rho_\mathcal{E} = \frac{1}{2^N}\hat{\mathcal{I}}$,
the bath-spin distribution is $\lambda_{I_{\mathcal{E}}} \approx
{I^2_{\mathcal{E}}}\exp(-{I^2_{\mathcal{E}}/2N})$. For this state $\langle \hat{I}^2_{\mathcal{E}} \rangle \sim N$.

The dynamical evolution of the system is governed by the equation $\rho (t) = U \rho (0) U^{\dagger}$, where $U$ is the time evolution operator. 
The two-qubit reduced density matrix $\rho_{AB}(t)$ is obtained through a partial trace over
the bath degrees of freedom. 
$\rho_{AB}(t)$ can in general be a mixed state, which would then imply that Tr$\rho^2_{AB}(t) < 1$.  We can use the extent of mixing as a measure of decoherence which is given by
\begin{eqnarray}
\label{decmeas}
{D}(t) &\equiv & 1-\text{Tr}\rho^2_{AB}(t)  \nonumber \\
&=& \frac{1}{4}[3-(P^2_A(t)+P^2_B(t))+\sum_{mn}(\Pi^{mn})^2(t))].
\end{eqnarray}
The above measure has a minimum value of zero for a pure state and a maximum value of $\frac{3}{4}$ for the completely
mixed state ($\rho_{AB} = \frac{1}{4}\mathcal{I}$). As a measure of entanglement, we shall use the concurrence measure 
\cite{woott} in our analysis. 
For a pure state, the concurrence is simply given by $C = 1-P_A^2$. Even though the expression for the concurrence
for a general mixed state is
not that straightforward, it can be expressed in terms of the $P^n_{A,B}$ and $\Pi^{mn}$. For example
if $[\rho_{AB}, S^z_A+S^z_B]=0$, then the concurrence is given by
\begin{eqnarray}
C &=& \frac{1}{2}{max}\bigg\lbrace\sqrt{(\Pi^{xx}+\Pi^{yy})^2+(\Pi^{xy}-\Pi^{yx})^2} \nonumber \\
&& -\sqrt{(1+\Pi^{zz})^2+(P^z_A+P^z_B)^2},0\bigg\rbrace.
\end{eqnarray}

In the next section, we study the dynamics for non-interacting qubits. We determine the evolution of the
the reduced density matrix $\rho_{AB}$ analytically and thereby extract the decoherence measure $D(t)$. In Sec-III we consider the case of interacting qubits ($J\ne 0$). 
The dynamics becomes more complicated, and is sensitive to whether the couplings of the qubits with the bath are same or different. 
We study both these cases, and the results are again obtained analytically. A short-time analysis of decoherence is made,
which allows us to identify a class of states with large decoherence time scales. We evaluate the decoherence measure and
the concurrence for longer times in Sec-IIIA ($K_A=K_B$) and in Sec-IIIB ($K_A\ne K_B$).

\section{Dynamics with non interacting qubits}
In this section we consider the case of non interacting qubits, i.e., the overlap between the spatial wave functions of the qubits is negligible. Thus, we set $J=0$.
As argued in Sec.I, the nuclear environments for the two qubits are different ($\vec{I}_{\mathcal{E}_A} \ne \vec{I}_{\mathcal{E}_B}$).
The Hamiltonian is 
\begin{eqnarray}
\mathcal{H}=K_A\vec{S}_A \cdot \vec{I}_{\mathcal{E}_A} + K_B\vec{S}_B \cdot \vec{I}_{\mathcal{E}_B}.
\end{eqnarray}
Since the two qubits evolve independent of each other, an initial direct product state will evolve into a direct product
state for later times, i.e.
if $\rho_{AB}(0)=\rho_A(0)\otimes\rho_B(0)$, then we have $\rho_{AB}(t)=\rho_A(t)\otimes\rho_B(t)$.
Hence the dynamics is of interest only when the initial state is entangled, which we study below.

Following the method developed in \cite{durga}, we can write down the unitary operator, that evolves the qubit-bath system,
as
\begin{eqnarray}
U(t) = (p_A + q_A\vec{S}_A\cdot \vec{I}_{\mathcal{E}_A})\otimes(p_B + q_B\vec{S}_B\cdot \vec{I}_{\mathcal{E}_B}).
\end{eqnarray}
Here the time-dependent coefficients are given by $p_{A}(t)=\cos \Lambda_{A}t +iK_{A}\sin \Lambda_{A}t /2\Lambda_{A}$,
$q_{A}(t)=2iK_{A}\sin \Lambda_{A}t/\Lambda_{A}$,  where $2\Lambda_{A} = K_{A}(I_{\mathcal{E}_{A}}+1/2)$, and a similar
form for $p_B(t), q_B(t)$ of the qubit $B$.
Using the above time-evolution operator, we evaluate the time dependent polarizations of the two-qubit state, and we obtain 
\begin{eqnarray}
\vec{P}_A(t) &=& g_1(t)\vec{P}_A(0),~~ \vec {P}_B(t)= \tilde g_1(t) \vec {P}_B(0)\nonumber \\
\Pi^{mn}(t) &=& g_2(t)\Pi^{mn}(0).
\end{eqnarray}
The time-dependent coefficients are given by
\begin{widetext}
\begin{eqnarray}
\label{gcoeff}
g_1(t) &=& \sum_{I_{\mathcal{E}_A},I_{\mathcal{E}_B}} \lambda_{I_{\mathcal{E}_A}} \lambda_{I_{\mathcal{E}_B}}|p_A p_B|^2
\left\lbrace 1 - { I_{\mathcal{E}_B}(I_{\mathcal{E}_B}+1)\over 12}\left[ \left|{q_B\over p_B}\right|^2- 3 
 {I_{\mathcal{E}_A}(I_{\mathcal{E}_A}+1)\over I_{\mathcal{E}_B}(I_{\mathcal{E}_B}+1)} \left|{q_A\over p_A}\right|^2 +
{I_{\mathcal{E}_A}(I_{\mathcal{E}_A}+1)\over 4}\left|{q_A q_B\over p_A p_B}\right|^2\right]\right\rbrace,\nonumber\\
g_2(t) &=& \sum_{I_{\mathcal{E}_A},I_{\mathcal{E}_B}} \lambda_{I_{\mathcal{E}_A}} \lambda_{I_{\mathcal{E}_B}}|p_A p_B|^2
\left\lbrace 1 - { I_{\mathcal{E}_B}(I_{\mathcal{E}_B}+1)\over 12}\left[ \left|{q_B\over p_B}\right|^2+  
 {I_{\mathcal{E}_A}(I_{\mathcal{E}_A}+1)\over I_{\mathcal{E}_B}(I_{\mathcal{E}_B}+1)} \left|{q_A\over p_A}\right|^2 -
{I_{\mathcal{E}_A}(I_{\mathcal{E}_A}+1)\over 12}\left|{q_A q_B\over p_A p_B}\right|^2\right]\right\rbrace.
\end{eqnarray}
\end{widetext}
The coefficient $\tilde g_1(t)$ is obtained from $g_1(t)$ by interchanging
the labels $A$ and $B$.
We note that $g_1^2(t) \ge g_2^2(t)$ i.e., the initial tensor polarization $\Pi^{mn}$ decays faster in comparison to the vector polarization. 
The decoherence measure (given in Eq.\ref{decmeas}) for the present case has a simple form and is given by
\begin{eqnarray}
D(t) = \frac{1}{4}\lbrace 3(1-g^2_2(t))-2[g^2_1(t)-g^2_2(t)]P(0)^2\rbrace.
\end{eqnarray}
where $P(0)=P_A(0)=P_B(0)$.
Since the initial concurrence is given by $C(0) = 1-P(0)^2$, it is clear 
that decoherence is sensitive to the entanglement in the initial state.
Decoherence becomes stronger 
 with increasing initial entanglement. These results are shown in Fig.1 where we have
plotted $D(t)$ for states with different initial entanglement. We can now extract the time scales for decoherence and the 
entanglement loss from the short-time behavior of the above time-dependent coefficients. For this, we consider 
$K_A = K_B = K$, and also $\langle \hat{I}^2_{\mathcal{E}_A}\rangle = \langle \hat{I}^2_{\mathcal{E}_B} \rangle =\langle
\hat{I}^2_{\mathcal{E}} \rangle$ i.e., the individual baths have the same bath-spin distribution,
and each qubit is interacting with its bath with the same interaction strength. The leading-order time dependence of the
coefficients are given by
\begin{eqnarray}
g_1(t) &\approx& 1-\frac{1}{3}K^2\langle \hat{I}^2_{\mathcal{E}} \rangle t^2\nonumber \\
g_2(t) &\approx& 1-\frac{2}{3}K^2\langle \hat{I}^2_{\mathcal{E}}\rangle t^2
\end{eqnarray}
The decoherence measure for short times is then given by $D(t) \approx 1- \exp{-t^2/\tau^2_D}$, with the decoherence
time scale given by 
\begin{eqnarray}
\frac{1}{\tau^2_D} = \frac{1}{3}K^2\langle \hat{I}^2_{\mathcal{E}} \rangle(3-P(0)^2)
\end{eqnarray}

As can be seen from the above, any entanglement present in the initial two-qubit state would imply $P(0)<1$, which in turn
will decrease the decoherence time scale.
Thus, we conclude that initial separable two-qubit states decoher less compared to the entangled states when the
exchange interaction between the qubits is negligible, and the two qubits interact with different sets of nuclear spins. It should be noted that the
decoherence time scale depends on the initial state of the bath and the qubits, which is a hallmark of a non-Markovian
evolution \cite{durga}.
\begin{figure}[htb]
   \includegraphics[width=8.0cm]{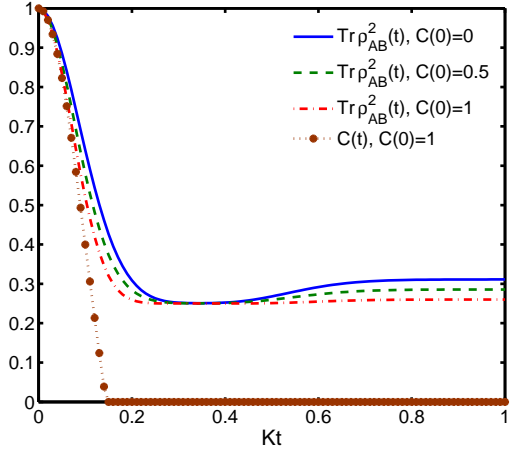}
 \caption{Purity measure $\text{Tr}[\rho^2_{AB}(t)]$ for a state with an
initial concurrence $C(0)$, is plotted as a function of time. The initial state of the baths are unpolarized consisting of $N=100$ spin $1/2$ nuclei.
It can be seen from the figure that, the loss of purity is faster for entangled states.
Also plotted here is the concurrence $C(t)$, for a maximally entangled state.
The loss of entanglement is faster than the loss of purity for a maximally entangled state.
We have taken $K_A = K_B = K$.}
 \end{figure}
For initial states with a nonzero entanglement, i.e. the concurrence at any
later time has a complicated structure.

However, for an initial two-qubit state $|\uparrow\downarrow+r \downarrow\uparrow>/\sqrt{1+r^2}$, which
is an eigenstate of $S^z$, the concurrence is straightforward to calculate
using Eqs.5, 8 and 9. The initial
polarizations here are given by $ P(0)=(1-r^2)/(1+r^2), \Pi^{zz}=-1,\Pi^{xx}=\Pi^{yy}=2r/(1+r^2)$. The short-time
behavior of the concurrence, using the short-time expansions of the time-dependent coefficients $g_1$ and $g_2$,
is $C(t) \approx C(0)\exp (-t^2/\tau^2_C)$, where the decay time scale is given by
\begin{eqnarray}
\frac{1}{\tau^2_C} ={1\over 3} K^2\langle \hat{I}^2_{\mathcal{E}}\rangle
{3-2P(0)^2\over 1 - P(0)^2}.
\end{eqnarray}
It should be noted that for initial unentangled states, the concurrence is zero for all times, as the evolution of the
two qubits is independent of each other, and the state is a direct product for all times.
It is clear from the above expressions for the time scales that $\tau_c\le \tau_D$ for the entangled states, and
the equality holds for the maximally-entangled states ($P(0)$=0 here).
Even though $C(t)$ and $D(t)$ have a similar short-time behavior, the concurrence falls off rapidly to zero for later times,as shown in Fig.1. The entanglement in the state vanishes when $g_2(t) \le 1/3$.

\section{Dynamics with interacting qubits}
In this section we consider the case of interacting qubits. This corresponds to a considerable overlap of
the spatial wave functions of the two qubits, implying a nonzero exchange interaction, and consequently the qubits see a
common nuclear bath ($\vec{I}_{\mathcal{E}_A}=\vec{I}_{\mathcal{E}_B}=\vec{I}_\mathcal{E}$). The exchange interaction is larger than 
the Overhauser field i.e., $J \gg (K_A+K_B)\sqrt{\langle \hat{I}^2_\mathcal{E} \rangle}$.
Thus we write the Hamiltonian as,
\begin{eqnarray}
\label{hamilsolv}
\mathcal{H}=(K_A\vec{S}_A+K_B\vec{S}_B)\cdot\vec{I}_\mathcal{E} + J\vec{S}_A \cdot \vec{S}_B.
\end{eqnarray}
The time-evolution operator in this case can be written as,
\begin{eqnarray}
\label{unitop}
U&=& \left[ a_1(t)+a_2(t)(\vec{S}_A-\vec{S}_B)\cdot \vec{I}_\mathcal{E} \right] (1-\frac{\hat{S}^2_{AB}}{2}) \nonumber \\
&& +\left [ a_3(t) + a_4(t)\vec{S}_{AB}\cdot \vec{I}_\mathcal{E} + a_5(t)(\vec{S}_{AB}\cdot \vec{I}_\mathcal{E})^2 \right. \nonumber \\
&& + \left. a_6(t)(\vec{S}_A-\vec{S}_B)\cdot \vec{I}_\mathcal{E} 
+ a_7(t)(\vec{S}_A \times \vec{S}_B)\cdot \vec{I}_\mathcal{E} \right ]  \frac{\hat{S}^2_{AB}}{2}. \nonumber \\
\end{eqnarray}
The expressions for the coefficients $a_i(t)$ are given in Appendix A (Eq.\ref{acoeff}).
Note that the coefficients $a_2(t), a_6(t)$ and $a_7(t)$ survive only when there is an asymmetry in the qubit bath coupling strengths ($K_A \ne K_B$). The operators $(\vec{S}_A-\vec{S}_B)\cdot \vec{I}_\mathcal{E} $ and $(\vec{S}_A \times \vec{S}_B)\cdot \vec{I}_\mathcal{E} $ appearing in
Eq.\ref{unitop} cause transitions between the singlet and triplet subspace of the qubits.

The time dependent polarizations are given by 
\begin{eqnarray}
\label{timdcoeff}
\vec{P}_A(t)&=& f_1(t)\vec{P}_A(0) + f_2(t)\vec{P}_B(0) \nonumber \\
            &&+f_3(t)\sum_{n=1}^{3}\hat{e}_n\sum_{i,j=1}^3\epsilon_{n i j}\Pi^{ij}(0),\nonumber \\
\vec{P}_B(t)&=& {f}_4(t)\vec{P}_B(0) + {f}_5(t)\vec{P}_A(0) \nonumber \\
&&- {f}_6(t)\sum_{n=1}^{3}\hat{e}_n\sum_{i,j=1}^3\epsilon_{n i j}\Pi^{ij}(0),\nonumber \\
\Pi^{mn}(t)&=& f_7(t)\Pi^{mn}(0) + f_8(t)\Pi^{nm}(0), \nonumber \\
&&+ f_9(t)\delta_{mn}\text{Tr}\Pi + f_{10}(t)\epsilon_{mnk}(P^k_A(0) - P^k_B(0)). \nonumber \\
\end{eqnarray}
where $\hat e_n$ stands for a cartesian unit vector. The expressions for the coefficients $f_i(t)$
are given in Appendix \ref{polcoeff}. 
We extract the decoherence time scale $\tau_D$ from the leading-order time dependence of the coefficients $f_i(t)$ for
short times, and we have 
\begin{eqnarray}
&&f_1(t) \approx 1-(\langle \hat{I}^2_{\mathcal{E}} \rangle K^2_A/3 + J^2/4)t^2, \nonumber \\
&&f_2(t),{f}_5(t) \approx J^2t^2/4, \nonumber \\
&&f_3(t),{f}_6(t), f_{10}(t) \approx Jt/2, \nonumber \\
&&f_4(t) \approx 1-(\langle \hat{I}^2_{\mathcal{E}} \rangle K^2_B/3 + J^2/4)t^2, \nonumber \\
&&f_7(t) \approx 1-\lbrace \langle \hat{I}^2_{\mathcal{E}} \rangle(K^2_A+K^2_B)/3+J^2/4 \rbrace t^2, \nonumber \\
&&f_8(t) \approx -(\langle \hat{I}^2_{\mathcal{E}} \rangle K_A K_B/3-J^2/4)t^2, \nonumber \\
&&f_9(t) \approx \langle \hat{I}^2_{\mathcal{E}} \rangle K_A K_B/3 t^2. \nonumber \\
\end{eqnarray}
Using the above short-time behaviour of the coefficients, we have 
$D(t) \approx 1-\text{e}^{-(\frac{t}{\tau_D})^2}$, and
the decoherence time scale $\tau_D$ is given by
\begin{eqnarray}
\label{tauexp}
\frac{1}{\tau^2_D}&=&\frac{1}{6}\langle \hat{I}^2_{\mathcal{E}} \rangle \left[(K^2_A+K^2_B)(3-P^2(0))\right. \nonumber \\
&&\left.+K_AK_B(\text{Tr}\Pi^2(0) - (\text{Tr}\Pi(0))^2)\right].
\end{eqnarray}
Let us recall that $P(0) = P_A(0) = P_B(0)$ for pure two-qubit states. If the initial state is separable,
then we further have, $\text{Tr}\Pi^2(0) = (\text{Tr}\Pi(0))^2$. The
decoherence time scale for these states is given by
\begin{eqnarray}
\label{septau}
\left ( \frac{1}{\tau^2_D} \right)_{sep} = \frac{1}{3}\langle \hat{I}^2_{\mathcal{E}} \rangle(K^2_A+K^2_B) \equiv \frac{1}{\tau^2_A}+\frac{1}{\tau^2_B},
\end{eqnarray}
where $\tau_A$ and $\tau_B$ represent the decoherence time scales for the individual qubits. Clearly
the individual qubits decoher slower in comparison to the two-qubit separable
state.
For maximally-entangled initial states ($P(0)=0$), the decoherence time scale depends on 
$\mathcal{R}=\text{Tr}\Pi^2(0) - (\text{Tr}\Pi(0))^2$, which further depends on the details of the state. For example, 
$\mathcal{R}=2$ for the Bell states in the triplet sector, and $\mathcal{R}=-6$ for the singlet state.
For the two cases, the decoherence time scales are respectively given by, 
\begin{eqnarray}
\label{maxenttau}
\left ( \frac{1}{\tau^2_D} \right)_{S} &=& \frac{1}{2}\langle \hat{I}^2_{\mathcal{E}} \rangle(K_A-K_B)^2, \nonumber \\
\left ( \frac{1}{\tau^2_D} \right)_{T} &=& \frac{1}{2}\langle \hat{I}^2_{\mathcal{E}} \rangle(K^2_A+K^2_B+
\frac{2}{3}K_AK_B).\nonumber \\
\end{eqnarray}
The time scales are sensitive to the relative sign between $K_A$ and $K_B$. If $K_A$ and $K_B$ both have a positive sign
(antiferromagnetic interaction) or a negative sign (ferromagnetic interaction), then the 
triplet sector suffers a faster decoherence, while the singlet state decohers faster if they have opposite signs.
It can also be
seen from Eq.\ref{septau} and Eq.\ref{maxenttau} that the separable states decoher slower than the entangled states
if $K_A$ and $K_B$ have opposite signs. For partially-entangled states, the calculation becomes more involved, and a
global minimization over the initial two-qubit pure states for finding the least decohering states will be done in Sec-IIID.
For the maximally-entangled Bell states ($C(0)=1$), the concurrence for later times is given by $C(t) = {max}\lbrace |\Pi^{xx}(t)| -
|1+\Pi^{zz}(t)|/2, 0\rbrace$. It has a
small-time behavior $C(t) \approx {\rm e}^{-(t/\tau^2_C)}$, with a decay time scale equal to the decoherence time scale
given above for the singlet and triplet state. Thus, the loss of entanglement and decoherence occur over the same time
scale. This feature for the maximally-entangled states, that $\tau_C=\tau_D$, we have seen in the case of the qubits
interacting with different nuclear baths also, except that there all the bell states have the same decoherence time scale. 
Below we examine the dynamics for larger times, $t>\tau_D$ in detail, with symmetric and asymmetric couplings separately.

\subsection{Identical coupling strengths $K_A = K_B = K$}
When $K_A=K_B$, the dynamics becomes simple, as the triplet and the singlet
components of a two-qubit state evolve independently.
The singlet sector is unaffected, and a nontrivial evolution occurs only in the triplet subspace. 
The time-evolution operator given in Eq.\ref{unitop} simplifies now to
\begin{eqnarray}
U &=& (1-\frac{\hat{S}^2_{AB}}{2}) +[a_3(t)+a_4(t)\vec{S}_{AB}\cdot
\vec{I_{\mathcal{E}}} \nonumber \\
&&+ a_5(t)(\vec{S}_{AB}\cdot \vec{I_{\mathcal{E}}})^2]\frac{\hat{S}^2_{AB}}{2}. 
\end{eqnarray}
We now consider the evolution of the  initial two-qubit states either maximally-entangled or unentangled separately. 
\subsubsection{Unentangled states}
For the unentangled state, let $\rho_{AB}(0) = | \uparrow \downarrow \rangle \langle \uparrow 
\downarrow |$. The initial polarizations of this state are $P^z_A=-P^z_B = 1$ and $\Pi^{zz}=-1$. 
With time, the system picks up other components of tensor polarization. The 
nonzero polarizations and the concurrence at any later time $t$ are given by
\begin{eqnarray}
P^z_A(t)&=&-P^z_B(t) = [f_1(t)-f_2(t)]P^z_A(0) \nonumber \\
\Pi^{zz}(t)&=&[f_7(t)+f_8(t)+f_9(t)]\Pi^{zz}(0) \nonumber \\
\Pi^{xx}(t)&=&\Pi^{yy}(t) =f_9(t)\Pi^{zz}(0) \nonumber \\
\Pi^{xy}(t)&=&-\Pi^{yx}(t)=f_{10}(t)[P^z_A(0)-P^z_B(0)]\nonumber \\
C(t)&=&\frac{1}{2}{max}\bigg\lbrace 2\sqrt{f^2_9(t)+f^2_{10}(t)} 
\nonumber \\
&&-|1-f_7(t)-f_8(t)-f_9(t)|,0\bigg\rbrace,
\end{eqnarray}
where the coefficients $f_i(t)$ are given in Appendix B.
In Fig.2 we have plotted the polarizations $P^z_A(t)$, $\Pi^{xx}(t)$ and $
\Pi^{zz}(t)$ as a function of time.
We have also plotted the concurrence in the same figure. One
can see that $P^z_A$ shows an under-damped behavior with a saturation value oscillating between
$\pm \frac{1}{3}$. The oscillation frequency is $2\pi/J$. The behavior of the
off-diagonal component of the tensor polarization $\Pi^{xy}$ is 
similar to that of $P^z_A$, but phase shifted by $\pi/2J$. 
The diagonal components of the tensor polarization contribute to the
concurrence, but have no dependence on $J$. 
The initial rapid oscillations in the concurrence are due to $\Pi^{xy}$. 
Hence, even when $J=0$, the two qubits get entangled. This is an example of bath-induced entanglement.

\begin{figure}[htb]
   \includegraphics[width=8.0cm]{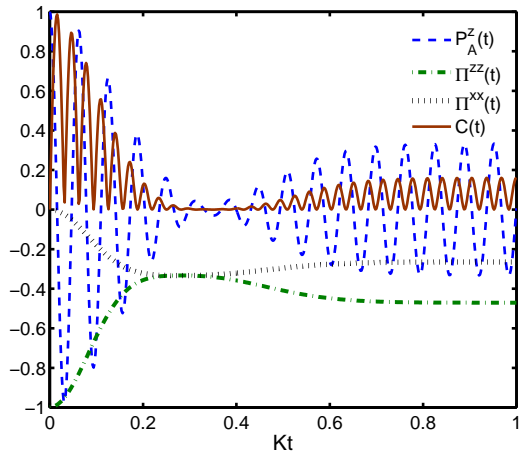}
 \caption{The vector and tensor polarizations are plotted as a function of time
for an unentangled state interacting with an initially unpolarized bath consisting of $N=100$ spin-$1/2$ nuclei. The time variation of concurrence $C(t)$ is also plotted.}
 \end{figure}

Since the initial two qubit-state is separable, the decoherence time scale can be calculated using Eq.\ref{septau}.
Taking the bath-spin distribution to be $\lambda_{I_{\mathcal{E}}}\approx$ $I_{\mathcal{E}}^2 \exp({-2I_{\mathcal{E}}^2/N})$,
we get $\tau_D = \frac{\sqrt{2}}{K\sqrt{N}}$.
As the individual qubits are also initially pure their corresponding decoherence time scale is given by $\tau_{A,B} = \frac{2}{K\sqrt{N}}$.
As argued above, the two-qubit state decohers faster in comparison to the individual qubits. Even though the two-qubit
decoherence is unaffected by the exchange interaction, the concurrence and the polarizations of individual qubits
are sensitive to $J$. 
A comparison between the single-qubit and the two-qubit decoherence measure is made in Fig.3

\begin{figure}[htb]
   \includegraphics[width=8.0cm]{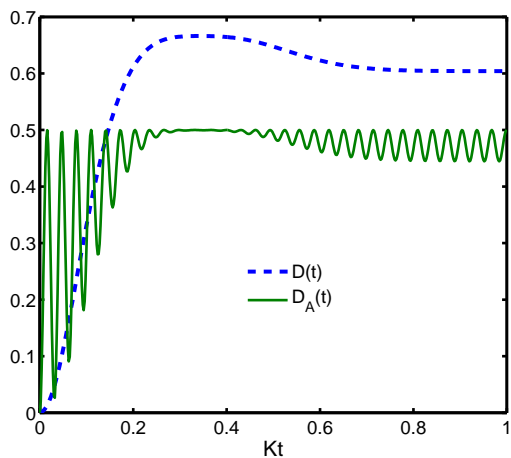}
 \caption{A comparison between the decoherence measure of the two qubit state $D(t)$, and that of the individual qubit
states
$D_A(t)=D_B(t)\equiv 1-\text{Tr}[\rho^2_{A}(t)]$, is shown as a function of time. The initial state of the
two qubits is unentangled, implying that the individual qubits are in a pure state initially.
 The initial state of the bath is unpolarized consisting of $N=100$ spin $1/2$ nuclei. }
 \end{figure}

\subsubsection{Maximally-entangled states}

\begin{figure}[htb]
   \includegraphics[width=8.0cm]{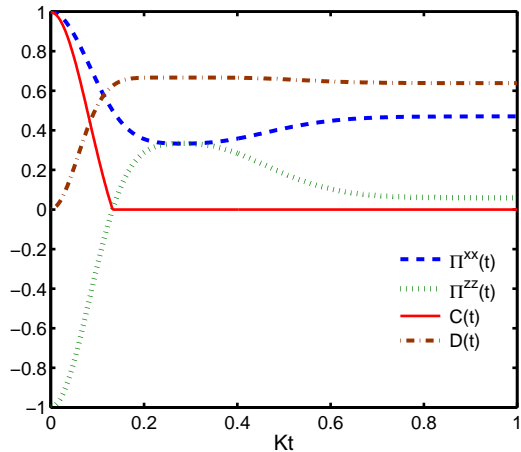}
 \caption{The non-vanishing components of tensor polarizations, the concurrence $C(t)$ and decoherence measure $D(t)$ are plotted as a function of time. The initial state of the two qubits is maximally entangled, and the bath is unpolarized. The bath consists of $N=100$ spin-$1/2$ nuclei.}
 \end{figure}
We shall now consider a maximally-entangled Bell state for the two-qubit state, viz. the triplet state with $S^z=0$.
The initial polarizations of this state are $\Pi^{xx}=\Pi^{yy} = 1$ and $\Pi^{zz}=-1$. The non-vanishing polarizations and
the concurrence at any later time $t$ are given by
\begin{eqnarray}
\Pi^{xx}(t)&=&\Pi^{yy}(t) =[f_7(t)+f_8(t)+f_9(t)]\Pi^{xx}(0) \nonumber \\
\Pi^{zz}(t)&=&[f_7(t)+f_8(t)-f_9(t)]\Pi^{zz}(0)\nonumber \\
C(t)&=&\frac{1}{2}{max}\bigg\lbrace2|\Pi^{xx}(t)|-|1-\Pi^{zz}(t)|,0 \bigg\rbrace
\end{eqnarray}
where $f_i(t)$ are given in Appendix B.
In Fig.4 we have plotted the components of tensor polarization, the concurrence and $D(t)$ as functions.
Setting $K_A = K_B = K$ in Eq.\ref{maxenttau}, and  taking the bath-spin distribution to be 
$\lambda_{I_{\mathcal{E}}}\approx$ $I_{\mathcal{E}}^2 \exp({-2I_{\mathcal{E}}^2/N})$, we have the decoherence
time scale $\tau_D = \frac{3}{K\sqrt{N}}$. For longer times $C(t)$ is zero, and at the instant $C(t)$ becomes zero,
the components of the tensor polarizations take values $\Pi^{xx} = \Pi^{yy} = 1/2$ and
$\Pi^{zz} = 0$.

In the case of non interacting qubits we have seen that all the components of the tensor polarizations had the same 
decay time scale.
In the present case the transverse components of tensor polarization ($\Pi^{xx}, \Pi^{yy}$) have different decay rates
in comparison to
the longitudinal component $\Pi^{zz}$. This can be seen from the small-time expansion of these polarizations which are given by 
\begin{eqnarray}
\Pi^{xx}(t) &=& \Pi^{yy}(t) = [1-2K^2\langle \hat{I}^2_\mathcal{E} \rangle t^2]\Pi^{xx}(0) \nonumber \\
\Pi^{zz}(t) &=& [1-4K^2 \langle \hat{I}^2_\mathcal{E} \rangle t^2]\Pi^{zz}(0).
\end{eqnarray} 
Denoting the longitudinal and transverse decay rates
by $\tau_1$ and $\tau_2$, we have $\tau_1 = 2\tau_2$. In contrast for the case of the noninteracting qubits, we have
$\tau_1=\tau_2$.

\subsection{Asymmetric couplings $K_A \ne K_B$}
In this section we consider the interaction (Eq.\ref{hamilsolv}) in its full generality. 
Here, the qubit-bath interaction term does not commute with the exchange term. 
This causes transitions between the singlet and triplet subspaces, due to which the singlet state also decohers. 
In Sec.III, we have seen that short-time dynamics is independent of $J$, for example, the decoherence time scale does not
depend on $J$. 
However, the effects of a nonzero $J$ become dominant for times $t > 1/J$, which can only be seen from the full solution
of the dynamics.
For $J$ large, the singlet and the triplet sectors of the qubits are
effectively  decoupled from each other. Since the hyperfine coupling strengths
are small compared to $J$, the transition matrix elements between the singlet
and
the triplet sectors are negligible. Thus, the singlet state and the states
with a large singlet fraction remain invariant through time evolution. On the
other hand, the triplet states suffer decoherence, due to transitions within
the triplet sector. Hence, the situations where a triplet is a ground state
or a metastable state are not favorable for quantum computations. 

We consider a one-parameter class of initial two-qubit states of the form
\begin{equation}
\label{onepstate}
|\psi_{AB}\rangle = \frac{1}{2(1+r^2)}[(1+r)|S_0\rangle + (1-r)|T_0\rangle],
\end{equation}
in terms of the singlet and triplet (with $S^z=0$) Bell states, i.e.
$|S_0\rangle = {1\over \sqrt{2}}[|\uparrow \downarrow \rangle - |\downarrow \uparrow \rangle]$
and $|T_0\rangle = \frac{1}{\sqrt{2}}[|\uparrow \downarrow \rangle + |\downarrow \uparrow \rangle]$. 
The evolution of the system is best studied in the Bell-basis, which apart from $|S_0\rangle$ and $|T_0\rangle$
also comprises of, $ |T_1\rangle = \frac{1}{\sqrt{2}}[|\uparrow \uparrow \rangle + |\downarrow \downarrow \rangle]$, $
|T_2\rangle = \frac{1}{\sqrt{2}}[|\uparrow \uparrow \rangle - |\downarrow \downarrow \rangle]$.
The two-qubit reduced density matrix state at any later time is given by
\begin{eqnarray}
\label{inhomostat}
\rho_{AB}(t) &=& c_1(t)|S_0\rangle \langle S_0| + c_2(t)|T_0\rangle \langle T_0| + c_3(t) T_0\rangle \langle S_0| 
\nonumber \\
&+&c_4(t)
\lbrace |T_1\rangle \langle T_1| + |T_2\rangle \langle T_2|\rbrace  + c_5(t) |T_1\rangle \langle T_2|\nonumber \\&+& H.C. 
\end{eqnarray}
The time-dependent coefficients are given by,
\begin{eqnarray} 
c_1(t) &=& \frac{1}{2(1+r^2)}\sum_{I_{\mathcal{E}}}\lambda_{I_{\mathcal{E}}}[(1+r)^2|a_1|^2 
+(1-r)^2 \nonumber \\
&&I_{\mathcal{E}}(I_{\mathcal{E}}+1) (|a_6|^2+|a_7|^2+2\text{Im}~ a^{*}_6a_7], \nonumber \\
c_2(t) &=&\frac{1}{2(1+r^2)}\sum_{I_{\mathcal{E}}}\lambda_{I_{\mathcal{E}}}[(1+r)^2(1-|a_1|^2)/3 \nonumber \\ &&+{(1-r)^2}\lbrace|a_3|^2+\frac{1}{3}I_{\mathcal{E}}(I_{\mathcal{E}}+1) \nonumber \\
&&((8I^2_{\mathcal{E}}+8I_{\mathcal{E}}-1){|a_5|^2\over5}+2\text{Re}~a_3 a^{*}_5)\rbrace],\nonumber 
\end{eqnarray}
\begin{eqnarray}
c_3(t)&=& \frac{1-r^2}{2(1+r^2)}\sum_{I_{\mathcal{E}}}\lambda_{I_{\mathcal{E}}}[a^{*}_1(a_3 + \frac{2}{3}I_{\mathcal{E}}(I_{\mathcal{E}}+1)a_5)\nonumber \\ &&+ \frac{1}{3}I_{\mathcal{E}}(I_{\mathcal{E}}+1)(a_2(a_6^{*}+ia^{*}_7)],\nonumber \\
c_4(t)&=& \frac{1}{2}[1-c_1(t)-c_2(t)], \nonumber \\
c_5(t) &=&\frac{r^2-1}{6(1+r^2)}\sum_{I_{\mathcal{E}}}\lambda_{I_{\mathcal{E}}}I_{\mathcal{E}}(I_{\mathcal{E}}+1)\text{\text{Re}}[a^{*}_2(a_5+2a_4)],\nonumber \\
\end{eqnarray}
where the coefficients $a_i$ carry time dependence, and are given in the Appendix A. If the initial state is one of the Bell states, $\rho_{AB}(t)$ will 
be diagonal in the Bell-basis for all times, because the coefficients $c_3(t) = 
c_4(t) =0$. For a general linear superposition state, $\rho_{AB}(t)$ picks up
off-diagonal terms as well.

To show the effects
of $J$ on the states close to the triplet and singlet states, we consider
two states corresponding to $r=\pm 0.5$ in Eq.25. 
The state corresponding to $r=-0.5$, is close to the triplet state, whereas
the other state with $r=0.5$ is close to the singlet. As argued earlier, we should expect a large dependence of $D(t)$
on $J$ for the state close to the singlet. This, in fact, can be seen from
Fig.5,
where we have plotted $D(t)$ for these states for two values of $J$. The decoherence measure $D(t)$ for the case
of $r=0.5$, is suppressed by a strong exchange interaction, whereas for the former case ($r=-0.5$), there is only a small
effect of $J$ on $D(t)$.
We note that for $r \rightarrow 1$ (singlet state) $D(t) \rightarrow 0$ for large values of $J$.
In the other limit when $r \rightarrow -1$ (triplet state), decoherence  persists and remains unaffected by the value of $J$.

As we have seen earlier in Sec-III, that up to $\large{O}(t^2)$ there is no dependence of $D(t)$ on $J$.
But when $J$ is large the contributions from higher-order terms also become significant over time scales $\tau < 1/K$.
\begin{figure}[htb]
   \includegraphics[width=8.0cm]{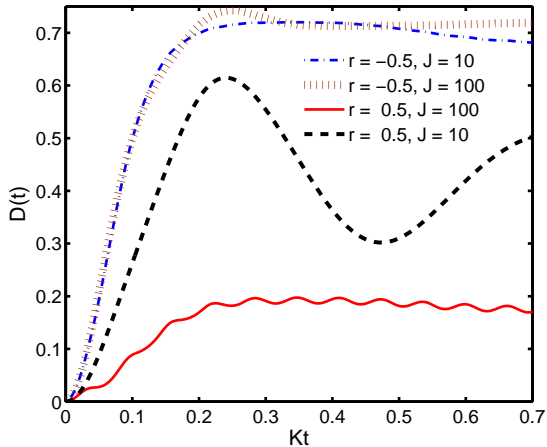}
 \caption{A plot of $D(t)$ for the partially entangled states corresponding to $r = \pm0.5$ (see Eq.\ref{onepstate})  is shown for different values of the exchange interaction strength $J$. The initial state of the bath is unpolarized consisting of $N
=100$ spin $1/2$ particles. Here $K = (K_A+K_B)/2$ is the average of the qubit bath interaction strengths.}
 \end{figure}
The initial Gaussian decay is seen only for times $t < 1/J$, and the behavior can change drastically for later times. It can be seen from Fig.5, that the behavior of $D(t)$ for the state corresponding to $r=0.5$ is similar (Gaussian) initially $(t < 1/J)$ and then changes into a slowly varying oscillatory function.

To understand the behavior of $D(t)$ for large $J$, we consider the singlet state, whose decoherence is highly sensitive to $J$. As can be seen from Eq.\ref{inhomostat}, the time evolved state has a simple structure, we have 
$\rho_{AB}(t) = c_1(t)|S_0\rangle \langle S_0|+(1-c_1(t))[|T_0\rangle \langle T_0|+
|T_1\rangle \langle T_1| + |T_2\rangle \langle T_2|]/3$, where $c_1(t)$ is given by
\begin{eqnarray}
\label{powerlaw}
c_1(t) = \sum_{I_\mathcal{E}}\lambda_{I_{\mathcal{E}}}[ \cos^2{\Lambda_{-}t} + p^2 \sin^2{\Lambda_{-}t}].
\end{eqnarray}
The coefficients $\Lambda_{-}$ and $p$ are given in Appendix A. For $J \gg \sqrt{N}(K_A+K_B)$, we can
approximate the coefficients $\Lambda_{-}$ and $p$ as
\begin{eqnarray}
\Lambda_{-} &\approx& \frac{J}{2}\left [1+\frac{(K_A-K_B)^2}{2J^2}I^2_{\mathcal{E}} \right] \nonumber \\
p^2 & \approx & 1-2\frac{(K_A-K_B)^2}{2J^2}I^2_{\mathcal{E}}
\end{eqnarray}
For a bath-spin distribution given by $\lambda_{I_{\mathcal{E}}}\approx$ $I_{\mathcal{E}}^2 \exp({-2I_{\mathcal{E}}^2/N})$, we find
\begin{eqnarray}
c_1(t) = 1-3\beta^2 \left [ 1-\frac {\cos(Jt+\frac{5}{2} \tan^{-1}{\beta t} ) } { (1+\beta^2 t^2) ^ {\frac{5}{4}}} \right ],
\end{eqnarray}
where $\beta = (K_A-K_B)\sqrt{N}/2J$. Here the decoherence measure is given by $D(t) = 2/3(1-c^2_1(t))$.
For large $J$, since $c_1(t) \approx 1$, $D(t)$ is quite small, implying that the singlet state hardly decohers. As can be seen from the above, there is a slow-varying oscillatory behavior for the decoherence measure. A similar power-law behavior
for the singlet-triplet correlator for large $J$ was shown earlier by Coish and Loss \cite{loss}. 
This also explains the oscillatory behavior of $D(t)$ for large $J$ for the state corresponding to $r=0.5$ (which is close to the singlet) seen in Fig 5.

The Bell states, which are widely used in quantum information protocols, rather have a special evolution. As noted earlier these states remain diagonal in the Bell basis for later times. For a Bell state we have $\vec P_A =\vec P_B =0$, and only the 
diagonal components of $\Pi$ to be non vanishing. The state at any later time has the form
\begin{eqnarray}
\rho_{AB}(t) = \frac{1}{4}[\mathcal{I}+4\sum_m \Pi^{mm}(t){S}^m_A {S}^m_B].
\end{eqnarray}
It follows from Eq.\ref{timdcoeff}, that $\Pi^{mm}(t) = [f_7(t)+f_8(t)]\Pi^{mm}(0)+f_9(t)\text{Tr}\Pi$. This is true only when the bath is unpolarized. The state can pick up other polarizations if there is a nonzero magnetic field, or when the bath
is polarized. We do not consider such cases in this work.

Before we end this section we consider a class of maximally-entangled mixed states, viz. the Werner states. A Werner state
is an admixture of the singlet and the identity state
viz., $\rho_{AB} =p |S_0\rangle \langle S_0 | + \frac{1-p}{4}\mathcal{I}$.
where $ |S_0\rangle $ is the two-qubit singlet state. This state remains invariant
under time evolution for all values of $p$ ($0 \le p\le 1$).  The concurrence for the above given state $C = \rm max [(3p-1)/2,0]$. Werner states
are entangled for $p > 1/3$ and are separable for $p < 1/3$. 
When $K_A = K_B$ the Werner state remains unchanged as the singlet has no evolution. However, when $K_A \ne K_B$,
the state evolves as the singlet decohers. The bath effects on this state can drastically get reduced when $J \gg (K_A+K_B)\sqrt{\langle \hat{I}^2_\mathcal{E} \rangle}$.

\subsection{Least-decohered two-qubit pure states}

We now address the problem of identifying the least decohering two-qubit states through the time evolution. The initial
state of the system is a direct product of the two-qubit pure state (which may or may not be a direct product) and
the bath state (which is a direct sum over bath states from various spin sectors with a bath-spin distribution).
We make no assumption on the relative magnitudes or signs of $K_A$, $K_B$ and $J$.  The general expression for the decoherence time has been obtained earlier from the short-time expansions of the time-dependent coefficients (see Eqn.\ref{tauexp}).
The decoherence time scale for a two-qubit state is given by
\begin{eqnarray}
\frac{1}{\tau^2_D}& =& \frac{2}{3}\langle \hat{I}^2_{\mathcal{E}} \rangle \lbrace \langle(K_A\vec{S}_A+K_B\vec{S}_B)^2\rangle
-\langle K_A\vec{S}_A+K_B\vec{S}_B\rangle^2 \rbrace, \nonumber \\
\end{eqnarray} 
where the expectation values are evaluated in the initial direct product state of the qubits and the bath.
It should be noted that the decoherence time scale is independent of $J$, as the leading-order contribution for small $t$
is independent of $J$ for all the time-dependent coefficients.

We wish to minimize the above expression over all the pure states. To that end, we write the most general pure two-qubit
state as
\begin{eqnarray}
|\psi\rangle = {1\over \sqrt{1+|\gamma|^2}}[|\uparrow_{\hat{n}_1}\downarrow_{\hat{n}_2}\rangle-\gamma|\downarrow_{\hat{n}_1}
\uparrow_{\hat{n}_2}\rangle ],
\end{eqnarray}
where ${\hat{n}_1}$, ${\hat{n}_2}$ are two arbitrary directions in space. Here, $\gamma$ is any complex number which 
determines the entanglement. The ket $|\uparrow_{\hat{n}_1}\rangle$ is an
eigenstate of the operator $\vec S_1.\hat{n}_1$ with an eigenvalue +1/2.
Since the Hamiltonian is isotropic, and the bath is unpolarized, we can choose ${\hat{n}_1}=\hat{z}$, without any loss of
generality. Let us denote the polar angles of the other arbitrary unit vector $
\hat {n}_2$ by
$(\theta, \phi)$. 
The decoherence time scale for the above state is given by
\begin{eqnarray}
\frac{1}{\tau^2_D} &=&\frac{1}{3}\langle \hat{I}^2_{\mathcal{E}} \rangle (K^2_A+K^2_B)\left[1+\frac{2|\gamma|^2}{(1+|\gamma|^2)^2
}(1-\delta\cos\theta)\right. \nonumber \\
&&\left.-2\delta\cos^2(\theta/2)\frac{Re(\gamma)}{1+|\gamma|^2}\right],
\end{eqnarray}
where we have introduced a hyperfine-inhomogeneity interaction parameter $\delta = \frac{2K_AK_B}{K^2_A+K^2_B}$.
The decoherence time scale is independent of $\phi$, and it
it is easy to check that the optimal value of the decoherence time is for $\theta=0$, i.e. $\hat{n}_2=\hat{z}$ in the
above state. It is also straightforward to check
that $\gamma$ is real for the least-decohered state. This would imply that it suffices
to optimize the decoherence time for a state of the form
$|\psi>=|\uparrow\downarrow -\gamma \downarrow\uparrow\rangle/(1+\gamma^2)$, with $\gamma$ real.
From the examples considered in the earlier sections, one would have expected the above form for the least decohering state.
We can choose the value of $\gamma$ by optimization of the decoherence time scale as a function of the coupling constants.
In Sec-IIIC we saw that states with $\gamma > 0$ (see Eq.\ref{onepstate}) have long decoherence times when $K_A, K_B$ are of the same sign i.e., $\delta > 0$. 
Maximizing the decoherence time scale, we find the optimal value of $\gamma$ to be,
\begin{eqnarray}
\label{taumax}
&&\gamma_{\rm opt} = \frac{(1-\delta )-\sqrt{1-2\delta}}{\delta}; \hspace{3mm} -1 \le \delta \le \frac{1}{2}, \nonumber \\
&& \gamma_{\rm opt} = 1;  \hspace{3mm} \frac{1}{2} \le \delta \le 1.\nonumber \\ 
\end{eqnarray}

For $\delta\ge 1/2$, the least-decohered state is the singlet state ($\gamma=1$). For $\delta=0$, which corresponds to
either $K_A$ or $K_B$ is zero, i.e. one of the qubits is decoupled from bath, the direct product states decoher slower
than the
entangled states. In Fig.6, we have plotted the decoherence time for the
maximally-entangled states (both for the singlet with $\gamma=1$ and the triplet with $\gamma=-1$) and the separable
state ($\gamma=0$) as $\delta$ is varied. Also, we have plotted the best possible decoherence time (note that the
state itself is determined by $\gamma_{opt}$).
The inset in Fig.6 shows the variation of $\gamma_{opt}$ as a function
of $\delta$. It can be seen from the figure that separable states have a larger decoherence time compared to the
maximally-entangled states over a large range, i.e. $\delta < 1/3$, however, the state with the largest decoherence
time has an intermediate value of entanglement. 

The above analysis can be generalized for a more general Hamiltonian 
\begin{eqnarray}
\mathcal{H} = \sum_{i}[K^{i}_A \vec{S}_A + K^{i}_B \vec{S}_B]\cdot \vec{I}_i + J\vec{S}_A\cdot\vec{S}_B,
\end{eqnarray}
\begin{figure}[htb]
   \includegraphics[width=8.0cm]{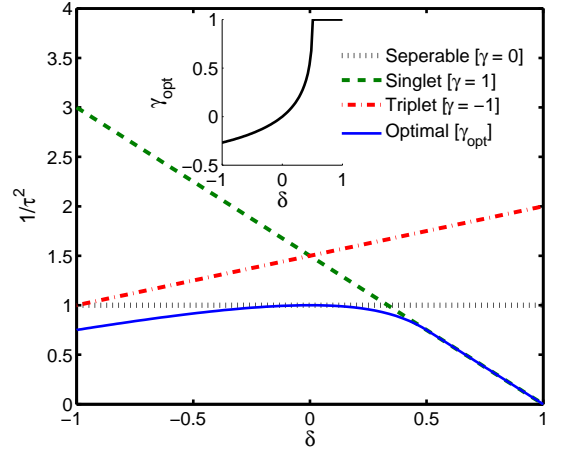}
 \caption{Variation of $1/\tau^2$ with $\delta$, for the separable state ($\gamma = 0$), the singlet ($\gamma = 1$) and the triplet ($\gamma = -1$) is shown above. Also shown is the dependence of $1/\tau^2$ on $\delta$,
for the least decohering state (Optimal). The values of $\gamma$ corresponding to the optimal curve are shown in the inset. The square of the inverse decoherence time ($1/\tau^2$) is plotted in units $K^2\langle\hat{I}^2_\mathcal{E}\rangle/3$.}
 \end{figure}

where the interaction strength between the qubit $A(B)$ and $i$'th nuclear-bath spin is denoted by $K_A(B)^i$. The
short-time analysis can easily be performed in this case aslo. 
The decoherence time scale for an initial two-qubit state $|\psi\rangle>=|\uparrow\downarrow-\gamma \downarrow\uparrow>/
(1+|\gamma|^2)$, interacting with $N$ nuclear spins, is given by
\begin{eqnarray}
\frac{1}{\tau^2_D} &=&{\eta_1\over 3} \langle \hat{I}^2_{\mathcal{E}}\rangle {\sum_i {K_A^i}^2 + {K_B^i}^2\over N}\left
 [ 1+\frac{2|\gamma|^2}{(1+|\gamma|^2)^2}(1-{\Delta}) \right .\nonumber \\
&&\left. - 2{\Delta}\frac{\rm \text{\text{Re}} (\gamma)}{1+|\gamma|^2} \right],
\end{eqnarray}
where ${\Delta} = \eta_2 \sum_i 2K^i_A K^i_B/(\sum_i {K_A^i}^2 + {K_B^i}^2)$ is the hyperfine-inhomogeneity parameter. We
have introduced two scale factors here $\eta_1$ and $\eta_2$ (both of which are of order unity), which depend on the
individual interactions strengths and also the bath-spin distribution. In general it is difficult to calculate the
scale factors. For an initial state of the bath of $N$ nuclear spins given by $\rho_{\mathcal{E}}= \frac{1}{2^N}\hat{\mathcal{I}}$, 
it is easy to calculate the scale factors, and we have $\eta_1=\eta_2=1$.
The value of $\gamma_{opt}$ for the
least-decohered state is given by Eq.35, except that the inhomogeneity parameter $\delta$ is replaced by $\Delta$. Taking the electron spacial wave functions
to be Gaussian functions, i.e. the groundstate wave functions of a 2-dimensional
confining harmonic oscillator potential appropriate to the quantum dots,
we find that the hyperfine-inhomogeneity parameter to be $\Delta\approx 0.6$,
when the distance between the qubits is of the order of the harmonic confinement
length. Thus, the case of a common nuclear bath considered in Sec-III holds 
for $\Delta>0.6$, as here $J\sim 0.1$ meV is larger than the hyperfine coupling
strengths. For $\Delta <0.6$, the common set of nuclear spins that interact with
the two qubits becomes smaller, and the qubits become isolated as $\Delta
\rightarrow 0$. In this case,  discussed in Sec-II, the exchange interaction
is negligible, and the two qubits evolve independent of each other, interacting
with two different sets of nuclear spins.

\section{Conclusion}

The dynamics of two spin-$1/2$ particles interacting with a nuclear spin bath in
a quantum dot is studied. The hyperfine interaction between
the electron spin and the nuclear spins is modeled by an isotropic Heisenberg interaction. The time-dependent polarizations
that characterize the two-electron spin (two-qubit) state are calculated as functions of the initial two-qubit state, and
the nuclear bath-spin distribution.
When there is negligible exchange interaction between the qubits themselves, corresponding to the situation where the two
qubits are physically apart, it is argued that each qubit sees a different set of nuclear spins as its environment. Here,
the separable states (i.e. unentangled two-qubit states) are found to have
have larger decoherence time scales in comparison to entangled states. On the other hand, when the exchange interaction
strength is appreciable, we argue that the two qubits interact with a common
nuclear spin bath. In this case, 
we have considered two cases:
the two qubits have same hyperfine coupling with the nuclear spins (symmetric coupling) and when the couplings are
different (asymmetric couplings). For maximally-entangled states,
the time scales corresponding to the loss of entanglement and decoherence are the same,
irrespective of whether the qubits have a common bath or not.
The effect of a nonzero exchange interaction strength $J$ comes is significant in the case of either there is an asymmetric
coupling between the qubits and the bath or 
the nuclear environment of the qubits are different. We have studied the effect of $J$ only in a limiting case, when 
the $J \gg (K_A+K_B)\sqrt{\langle \hat{I}^2_\mathcal{E}\rangle}$. For intermediate overlap of the electron wave functions, where
there are some nuclei in common and some different, analytical solution to dynamics is a much more difficult task. 
In such situations we have numerically seen that the dominant effect of $J$ arises when the initial state is either the
singlet or a state having large overlap with the singlet.
The least decohering two-qubit states are found by minimizing the decoherence time scale over all initial two-qubit pure
states, for a given hyperfine-inhomogeneity interaction.

\appendix
\section{\label{unitary}The unitary operator}
For the case of qubits interacting with a common nuclear bath, the Hamiltonian can be written as
\begin{eqnarray}
\mathcal{H}=(K_A\vec{S}_A+K_B\vec{S}_B)\cdot\vec{I}_\mathcal{E} + J\frac{\hat{S}_{AB}}{2},
\end{eqnarray}
where $S_{AB}$ is the total spin of the qubits, which can take two values 0 and 1. The total spin of the qubits and
the bath can take three values ${F}=I_{\mathcal E}-1,I_{\mathcal E},I_{\mathcal E}+1$. Denoting the projection
operator by $\hat P_{\mathcal{ J},S_{AB}}$ for a sector with a given total spin $F$ and the qubit spin $S_{AB}$,
we have
\begin{eqnarray}
\hat{P}_{I_\mathcal{E}+1, 1} &=& \frac{1}{(I_\mathcal{E}+1)(2I_\mathcal{E}+1)}[(\vec{S}_{AB}\cdot\vec{I}_\mathcal{E})^2 \nonumber \\
&&+(I_\mathcal{E}+2)(\vec{S}_{AB}\cdot\vec{I}_\mathcal{E}) + I_\mathcal{E}+1]\frac{\hat{S}^2_{AB}}{2}, \nonumber \\
\hat{P}_{I_\mathcal{E}-1, 1} &=& \frac{1}{I_\mathcal{E}(2I_\mathcal{E}+1)}[(\vec{S}_{AB}\cdot\vec{I}_\mathcal{E})^2 \nonumber \\
&&+(1-I_\mathcal{E})(\vec{S}_{AB}\cdot\vec{I}_\mathcal{E}) + I_\mathcal{E}]\frac{\hat{S}^2_{AB}}{2}, \nonumber \\
\hat{P}_{I_\mathcal{E}, 1} &=& 1-\frac{1}{I_\mathcal{E}(I_\mathcal{E}+1)}[(\vec{S}_{AB}\cdot\vec{I}_\mathcal{E})^2 
+\vec{S}_{AB}\cdot\vec{I}_\mathcal{E}]\frac{\hat{S}^2_{AB}}{2}, \nonumber \\
\hat{P}_{I_\mathcal{E}, 0} &=& (1-\frac{\hat{S}^2_{AB}}{2}).
\end{eqnarray}

There are two sectors with $F=I_{\mathcal {E}}$, which can mix if there is an asymmetry in the couplings
(i.e. $K_A\ne K_B$).
Since $F^z$ is conserved,
only states with the same $F^z$ can mix. In the subspace spanned by 
the two states with $F=I_\mathcal{E}$, the Hamiltonian matrix has the simple form
\begin{eqnarray}
\tilde{H} = \left[ \begin{array}{cc} {J}-\frac{K_A+K_B}{2} & \sqrt{I_\mathcal{E}(I_\mathcal{E}+1)}\frac{K_A-K_B}{2} \\
        \sqrt{I_\mathcal{E}(I_\mathcal{E}+1)}\frac{K_A-K_B}{2} & \ 0 \end{array} \right].
\end{eqnarray}
Since the matrix elements do not depend on the value of $F^z$, there are only two distinct eigenvalues 
corresponding to the two $F = I_\mathcal{E}$ sectors. The eigenvalues are given by
\begin{eqnarray}
\zeta_{\pm} &=& \frac{1}{2}(J-(K_A+K_B)/{2}) \nonumber \\
&&\pm\frac{1}{2}\sqrt{({J}-{(K_A+K_B)}/{2})^2+{I_\mathcal{E}(I_\mathcal{E}+1)}(K_A-K_B)^2}. \nonumber \\
\end{eqnarray}
The eigenvalues corresponding to the sectors with $F=I_{\mathcal {E}}+1,I_{\mathcal {E}}-1$ are given 
respectively as 
\begin{eqnarray}
\lambda_1 &=& {J}+I_\mathcal{E}{(K_A+K_B)}/{2}, \nonumber \\
\lambda_2 &=& {J}-(I_\mathcal{E}+1){(K_A+K_B)}/{2}. \nonumber \\
\end{eqnarray}
Now, the time-evolution operator in the full space can be
written as
\begin{eqnarray}
U& =& \text{e}^{i\lambda_1 t}\hat{P}_{I_\mathcal{E}+1, 1} + \text{e}^{i\lambda_2 t}\hat{P}_{I_\mathcal{E}-1, 1}
+\text{e}^{i\Lambda_{+}t}\lbrace \cos\Lambda_{-} t - i\sin{\Lambda_{-} t}\nonumber \\
&&[p(1+2\vec{S}_{AB}\cdot\vec{I}_\mathcal{E})+iq\frac{\vec{S}_A-\vec{S}_B}{\sqrt{I_\mathcal{E}(I_\mathcal{E}+1)}}]\rbrace (\hat{P}_{I_\mathcal{E}, 1} + \hat{P}_{I_\mathcal{E}, 1}) ,\nonumber \\
\end{eqnarray}
where
$\Lambda_{\pm} = (\zeta_{+} \pm \zeta_{-})/2$, $p = \frac{\Lambda_{+}}{\Lambda_{-}}$, and $q=\sqrt{1-p^2}$.
Thus, the unitary evolution operator can be written as
\begin{eqnarray}
U&=& \left[ a_1(t)+a_2(t)(\vec{S}_A-\vec{S}_B)\cdot \vec{I}_\mathcal{E} \right] (1-\frac{\hat{S}^2_{AB}}{2}) \nonumber \\
&& +\left [ a_3(t) + a_4(t)\vec{S}_{AB}\cdot \vec{I}_\mathcal{E} + a_5(t)(\vec{S}_{AB}\cdot \vec{I}_\mathcal{E})^2 \right. \nonumber \\
&& + \left. a_6(t)(\vec{S}_A-\vec{S}_B)\cdot \vec{I}_\mathcal{E} 
+ a_7(t)(\vec{S}_A \times \vec{S}_B)\cdot \vec{I}_\mathcal{E} \right ]  \frac{\hat{S}^2_{AB}}{2}. \nonumber \\
\end{eqnarray}
The time-dependent coefficients $a_i$ are given by
\begin{eqnarray}
\label{acoeff}
a_1(t) &=& e^{i\Lambda_{+}t}[\cos(\Lambda_{-}t)-ip\sin(\Lambda_{-}t)] ,\nonumber \\
a_2(t) &=& \frac{iqe^{i\Lambda_{+}t}}{\sqrt{I_{\mathcal{E}}(I_{\mathcal{E}}+1)}}\sin(\Lambda_{-}t), \nonumber \\
a_3(t) &=& a_1(t) + \frac{1}{2I_{\mathcal{E}}+1}(e^{i\lambda_1t}-e^{i\lambda_2t}), \nonumber \\
a_4(t) &=& \frac{1}{I_{\mathcal{E}}(I_{\mathcal{E}}+1)(2I_{\mathcal{E}}+1)}\left[I_{\mathcal{E}} (2+I_{\mathcal{E}}) e^{i\Lambda_{1}t} \right. \nonumber \\
&&\left. -({I}^2_{\mathcal{E}}-1)e^{i\lambda_2t} - a_1(t)(2I_{\mathcal{E}}+1) \right ],\nonumber \\
a_5(t) &=& \frac{1}{I_{\mathcal{E}}(I_{\mathcal{E}}+1)(2I_{\mathcal{E}}+1)}\left [I_{\mathcal{E}}e^{i\lambda_1t} 
\right. \nonumber \\
&&\left. +(I_{\mathcal{E}}+1)e^{i\lambda_2t} - a_1(t)(2I_{\mathcal{E}}+1) \right ]\nonumber, \\
a_6(t) &=& \frac{iqe^{i\Lambda_{+}t}}{\sqrt{(I_{\mathcal{E}}(I_{\mathcal{E}}+1))}}\sin(\Lambda_{-}t), \nonumber \\
a_7(t) &=& -\frac{qe^{\Lambda_{+}t}}{(I_{\mathcal{E}}(I_{\mathcal{E}}+1))^{\frac{3}{2}}}\sin(\Lambda_{-}t). \nonumber \\
\end{eqnarray}
For the simpler case of $K_A = K_B$, we have $\Lambda_+ = \Lambda_-$, $p=1$, $q=0$ implying that
$a_1(t) = 1$ and $a_2(t)=a_6(t)=a_7(t)=0$.
\section{\label{polcoeff}The time-dependent coefficients $f_i(t)$}
The time-dependent coefficients $f_i(t)$ (that determine the polarizations for any time $t$) occurring in Eq.16  have in
general a complicated structure. However, they simplify substantially for the case of $K_A=K_B$, which we give below
in terms of the time-dependent coefficients $a_i(t)$,
\begin{eqnarray}
f_0(t)&=&\sum_{I_\mathcal{E}}\lambda_{I_\mathcal{E}}\lbrace a_3+2a_5{ I_\mathcal{E}
(I_\mathcal{E}+1)\over 3} \rbrace, \nonumber \\
f_1(t) &=& f_4 = \frac{1}{4}\sum_{I_\mathcal{E}} \lambda_{I_\mathcal{E}}\lbrace 1+|a_3|^2 +[|a_5|^2 
\nonumber \\&& +4\text{Re} (a_3 
a^{*}_5) {I_\mathcal{E}(I_\mathcal{E}+1)\over 3}]\rbrace +{1\over 2}\text{ Re}f_0(t), \nonumber\\
f_2(t) &=& f_5(t) = f_1(t)-\text{Re} f_0(t),\nonumber \\
f_3(t) &=& f_6(t) = 2f_{10}(t) = \text{Im} f_0(t), \nonumber \\
f_7(t)&=&-\frac{1}{4}+\frac{3}{4}\sum_{I_\mathcal{E}}\lambda_{I_\mathcal{E}} \lbrace|a_3|^2+  [|a_5|^2
{(8I_\mathcal{E}(I_\mathcal{E}+1)-1)\over5} \nonumber  \\ && 
+ 4\text{Re}(a_3a^{*}_5)]{I_\mathcal{E} (I_\mathcal{E}+1)\over3}\rbrace + {1\over 2} \text{Re} f_0(t),
\nonumber \\
f_8(t) &=& f_7(t)- \text{Re} f_0(t), \nonumber \\
f_9(t) &=& \frac{1}{3}[1-(f_7(t)+f_8(t))].\nonumber \\
\end{eqnarray}

\end{document}